\documentclass[12pt]{article}
\usepackage{amsbsy,amssymb}

\usepackage[english]{babel}

\newcommand{\be}{\begin{eqnarray}}
\newcommand{\ee}{\end{eqnarray}}
\newcommand{\ben}{\begin{eqnarray*}}
\newcommand{\een}{\end{eqnarray*}}

\title{
On Dirac theory in the space with deformed Heisenberg algebra. Exact solutions}
\author{I. O. Vakarchuk\\
{\small Ivan Franko National University of Lviv}\\
{\small 12 Drahomanov Street, Lviv UA-79005, Ukraine}\\
{\small E-mail: }}
\date{ }
\begin{document}
\maketitle

\begin{abstract}
The Dirac equation has been studied in which  the Dirac matrices
$\hat{\mbox{\boldmath$\alpha$}}, \hat\beta$ have space factors,
respectively  $f$ and $f_1$, dependent on the particle's space
coordinates. The  $f$  function deforms Heisenberg algebra for
the coordinates and momenta operators, the function $f_1$ being
treated as a dependence of the particle mass on its position.
The properties of these functions in the transition to the
Schr\"odinger equation are discussed. The exact solution of  the
Dirac equation for the particle motion in the Coulomnb field with a
linear dependence of the  $f$  function on the distance  $r$  to
the force centre and the inverse dependence on  $r$  for the
$f_1$ function has been found.
\end{abstract}

\vspace{0.5cm}

\noindent {PACS numbers}:

\newpage

\section*{Introduction}
The problems with deformed Heisenberg algebra with small
additions to the canonical commutational relations have been
under a thorough and versatile scrutiny for a period of time
[1-10].  Deformed commutational relations were studied for
the first time in  \cite{11} where this issue was raised in
connection with the idea of quantisation of space. The question  of
deformation of the Heisenberg algebra can be approached along purely
practical lines when solving eigenvalues problems. When we have a Hamiltonian in the
Schr\"odinger equation with the potential not allowing to find the
exact analytical solution of the problem we can reduce it to  a
familiar form (for instance to the Hamiltonian of an harmonic
oscillator) using generalized coordinates and momenta that fail to satisfy the Heisenberg algebra. The
permutation relations between these operators are the so-called deformed
relations. With this procedure we transfer the
``inconvenient'' form of the Hamiltonian into a deformation of
Heisenberg algebra. Sometimes this procedure makes it possible to
more effectively find the approximate solutions of  the
Schr\"odinger equation. Some deforming functions allow to treat this
kind of transfer of inconveniences from the Hamiltonian
onto the permutational relations   as a problem in which the
particle mass is position-dependent.

We can start from the beginning with a ``good'' Hamiltonian
having deformed Heisenberg algebra with a certain arbitrary deforming
function dependent both on the coordinates and momentum.
Generally speaking, not always will we be able to make the inverse transition,
i.e. to ``toss'' this deformation back to the Hamiltonian.
For this matter  such problems are of interest in themselves  similarly to those about the
motion of a particle with a position-dependent mass. At the same, time we have to deal
with the problem of mutual ordering of the momentum operators and
the inverse mass in the kinetic energy. This problem, however,
does not appear when we resort to the Dirac equation.

Thus, we arrive at the possibility to formulate the problem about
the motion of the relativistic particle with a position-dependent mass
in the space with deformed Heisenberg algebra. To study this problem is the aim of this paper.
We also give the exact solution of the Dirac equation for the motion of a particle in the Coulomb
field when its mass and deforming function are specifically dependent on the coordinates.

\section{The initial equations}

Let us start from the Dirac equation for a particle   with the
potential energy
 $U$  in conventional notation:
\be
\left[(\hat{\mbox{\boldmath$\alpha$}}\hat{\bf
P})c+m^{*}c^2\hat{\beta}+U\right]\Psi=E\Psi, \label{Dir1} \ee
where $\hat{\mbox{\boldmath$\alpha$}}, \hat\beta$  are the Dirac
matrices, the coordinates and momenta satisfy
the permutational relations with deformed Heiselberg
algebra:
\be
\left\{
\begin{array}{l}
{\displaystyle [x_j,x_k]=0},\cr {\displaystyle [x_j,\hat
P_k]=i\hbar \delta_{jk}f},\cr {\displaystyle [\hat P_j,\hat
P_k]=-i\hbar\left({\partial f\over \partial x_j} \hat
P_k-{\partial f\over \partial x_k} \hat P_j\right), \ \
(j,k)=1,2,3};
\end{array}
\right. \label{spivv} \ee with the deforming function
$f=f(x,y,z)$ dependent on the particle coordimnates only. We
assume that the particle mass $m$ substituted for a certain
effective mass  $m^{*}$  is also position-dependent:
\be
m^{*}=m f_1,\ \ \ f_1=f_1(x,y,z). \label{mz} \ee
The embedding in the Dirac equation  of the functions  $f$ and $f_1$ implies
involvement  of  extra forces acting on the particle
alongside of those represented by the function $U$. We introduce a
new momentum:
\be
\label{imp} &&\hat{\bf p}=f^{-1/2}\hat{\bf P}f^{-1/2}, \nonumber\\
&&\hat{\bf P}=f^{1/2}\hat{\bf p}f^{1/2}, \ee in the way  that
coordinates and new momenta  become canonically conjugated
\be
\left\{
\begin{array}{l}
[x_j,x_k]=0,\cr [x_j,\hat p_k]=i\hbar \delta_{jk},\cr [\hat
p_j,\hat p_k]=0.
\end{array}
\right. \label{spryaz} \ee
Now the Dirac equation  (\ref{Dir1})
looks as follows:

\be
\left[f^{1/2}(\hat{\mbox{\boldmath$\alpha$}}\hat{\bf
p})f^{1/2}c+mc^2f_1\hat{\beta}+U\right]\Psi=E\Psi. \label{Dir2}
\ee
We make the transformation
\be
\bar \Psi=f^{1/2}\Psi, \label{peretv} \ee as a result of which
equation  (\ref{Dir2}) for the new function  $\bar \Psi$ will
be
\be
\left[f(\hat{\mbox{\boldmath$\alpha$}}\hat{\bf p})c
+mc^2f_1\hat{\beta}+U\right]\bar\Psi=E\bar\Psi. \label{Dir3} \ee

We can treat this equation as the  usual  Dirac equation in which
the Dirac matrices  $\hat{\mbox{\boldmath$\alpha$}}$ are multiplied
by certain position-dependent factors:
\be
\label{mnozh}
&&\hat{\mbox{\boldmath$\alpha$}}'=
f\hat{\mbox{\boldmath$\alpha$}}, \nonumber\\
&&\hat{\beta}'=f_1\hat \beta.
\ee

The matrix components $\hat{\mbox{\boldmath$\alpha$}}'
$ and the
matrix  $\hat\beta'$ are mutually  anticommuting.  The squares of
the components of the matrix $\hat{\mbox{\boldmath$\alpha$}}'$ equal
$f^2$, and the square of $\hat \beta'$ equals  $f_1^2$.

Before we consider the exact solutions of equation
(\ref{Dir3}) it is expedient to pass to the nonrelativistic
limit in the Dirac equation in order to find out the
properties of the functions $f$ and  $f_1$.

\section{The nonrelativistic limit. The Schr\"odinger equation}
\setcounter{equation}{0}

In order to receive the Schr\"odinger equation  from equation
(\ref{Dir3}) at $c\to \infty$ we introduce the new function
$\psi$ by the following relation:
\be
\bar \Psi=\left[f(\hat{\mbox{\boldmath$\alpha$}}\hat{\bf p})c
+mc^2f_1\hat{\beta}+E-U\right]\psi. \label{psi} \ee

After substituting  (\ref{psi})  in (\ref{Dir3}) we find:
$$\left\{{f(\hat{\mbox{\boldmath$\alpha$}}\hat{\bf
p})f(\hat{\mbox{\boldmath$\alpha$}}\hat{\bf p})\over 2m}
+{m^2c^4f_1^2-(E-U)^2\over 2mc^2}-{i\hbar f
(\hat{\mbox{\boldmath$\alpha$}}{\mbox{\boldmath$\nabla$}}U)\over
2mc} +{i\hbar c f\over 2}\hat
\beta\left(\hat{\mbox{\boldmath$\alpha$}}{\mbox{\boldmath$\nabla$}}f_1\right)
\right\}\psi=0.$$
We measure energy from the rest energy  $mc^2$, $$E'=E-mc^2,$$ and after simple transformations
we obtain:
\be
\label{def} &&\Bigg\{{f(\hat{\mbox{\boldmath$\alpha$}}\hat{\bf
p})f(\hat{\mbox{\boldmath$\alpha$}}\hat{\bf p})\over
2m}+U-{(E'-U)^2\over 2mc^2}-{i\hbar
f(\hat{\mbox{\boldmath$\alpha$}}\hat{\mbox{\boldmath$\nabla$}}U)\over
2mc} \nonumber\\ &&+{mc^2\over 2}\left(f_1^2-1\right)+{i\hbar c
f\over 2}\hat \beta\left(\hat{\mbox{\boldmath$\alpha$}}
{\mbox{\boldmath$\nabla$}}f_1\right)\Bigg\}\psi=E'\psi, \ee

From the latter two terms in the parentheses of equation
(\ref{def})  follows the condition on the behaviour of the
function  $f_1$ in the nonrelativistic limit. Indeed,
for the light velocity  $c$ to drop out of equation
(\ref{def}) when $c\to \infty$ it is necessary that  $f_1^2-1\sim
1/c^2$. The function $f_1$ can lead to one at  $c\to \infty$ also
faster than $1/c^2$ leaving no contribution whatsoever in the
nonrelativistic limit. If
\be
f_1^2-1={2\over mc^2}U_1,\ \ \ \ \ c\to \infty, \label{f1} \ee
where $U_1=U_1(x,y,z)$ is a certain function of the coordinates;
then from equation  (\ref{def}) we find its nonrelativistic
limit:
\ben \left[ {f(\hat{\mbox{\boldmath$\alpha$}}\hat{\bf
p})f(\hat{\mbox{\boldmath$\alpha$}}\hat{\bf p})\over 2m}
+U+U_1\right]\psi=E'\psi.
\een
We substitute $$\psi=\sqrt f\varphi,$$ and assuming that  the
function $f$ depends on the length  $r$ of the radius-vector
${\bf r}$  after simple transformations using the properties of the matrix
$\hat{\mbox{\boldmath$\alpha$}}$ we obtain  the following
equation:
\be
\left\{{(f^{1/2}\hat{\bf p}f^{1/2})^2\over 2m}+U+\Delta
U+U_1\right\}\varphi=E'\varphi, \label{f1/2} \ee
\be
\Delta U={f\over m r} {df\over d r}(\hat{\bf S}\hat{\bf L}),
\label{dU} \ee where $\hat{\bf S}={\hbar
\hat{\mbox{\boldmath$\sigma$}}/2}$  is the operator of the
particle spin, $\hat{\mbox{\boldmath$\sigma$}}=(\hat \sigma_x,\hat
\sigma_y,\hat \sigma_z)$ are the Pauli matrices, $\hat {\bf L}$  is
the angular momentum.

Expression (\ref{f1/2}) can be treated as the Schr\"odinger
equation for a particle with the position dependent mass  $\bar
m=m/f^2$ where the momentum operator and the inverse mass in the
kinetic energy operator are specifically ordered:
\be
\hat T={1\over \bar m^{1/4}}\, \hat{\bf p} {1\over \sqrt{\bar
m}}\,\hat {\bf p} {1\over \bar m^{1/4}}. \label{T} \ee
If the
particle has a spin then in the nonrelativistic limit  the
quantity $\Delta U$ remains. We refer to it as spin-orbital
deformation interaction.

If we write equation (\ref{f1/2}) using the ``old'' momentum
(\ref{imp}) we have the Schr\"odinger equation in the
space with deformed Heisenberg algebra:
\be
\left({\hat{\bf P}^2\over 2m}+U+\Delta
U+U_1\right)\varphi=E'\varphi. \label{Gajz} \ee
Hence, if in the
nonrelativistic theory we start  from the standard
Schr\"odinger equation for the study of the behaviour of the
particle with the deformed permutative relations  (\ref{spivv}),
the contribution from the spin-orbital interaction  $\Delta U$
gets lost  as well as  the term $U_1$ caused by the dependence of the particle mass on the coordinates
(1.3).

\section {The Dirac radial equation}
\setcounter{equation}{0}

We consider the particle motion in the central symmetrical field
$U$ and  functions $f,f_1$ to be dependent on the distance
$r$ only. We return to the Dirac equation  (\ref{Dir3}) and
reduce it to the radial equation. In order to do it we introduce
the radial momentum operator.
\be
\hat p_r=r^{-1}({\bf r}\hat{\bf p}-i\hbar) \label{radimp} \ee and
a radial component of the matrix
$\hat{\mbox{\boldmath$\alpha$}}$,
\be
\hat{\alpha}_r=(\hat{\mbox{\boldmath$\alpha$}}\hat{\bf n}),\ \ \ \
\ {\bf n}={{\bf r}\over r}. \label{rad} \ee

Further, following  \cite{12},we  introduce the  operator
\be
\hbar\hat K=\hat \beta
\left[(\hat{\mbox{\boldmath$\sigma'$}}\hat{\bf L})+\hbar\right],
\label{oper} \ee
$$
\hat{\mbox{\boldmath{$\sigma'$}}}=
\left(
\begin{array}{ll}
\hat{\mbox{\boldmath{$\sigma$}}} & 0\cr
0 & \hat{\mbox{\boldmath{$\sigma$}}}
\end{array}
\right),
$$
and calculating the product $\hat \alpha_r\hat K$ we transform equation (\ref{Dir3}) into the
following one:
\be
\left(f\hat \alpha_r\hat p_r c+{i\hbar c f\over r} \hat
\alpha_r\hat \beta\hat K+mc^2f_1 \hat \beta +U\right)\bar \Psi
=E\bar \Psi. \label{epsi} \ee
The operator $\hat K$ is the  motion
integral with the eigenvalues
\be
k=\pm \left(j+{1\over 2}\right)=\pm 1, \pm 2, \ldots\,,
\label{operat} \ee
$j$ is the quantum number of the total angular momentum. That is why in the
representation where the operator  $\hat K$
is diagonal the Dirac radial equation has the form:
\be
\left(f\hat \alpha_r\hat p_r c+{i\hbar c f\over r} \hat
\alpha_r\hat \beta k+mc^2f_1 \hat \beta +U-E\right)\bar R =0,
\label{barR} \ee and
\be
\bar \Psi=Y\bar R, \label{psir} \ee
$Y$ is the spherical spinor
that is the eigenvalue of the operator  $\hat K$, $\bar R$ is
the radial function. Now we introduce a new radial function
$R$  with the following relation:
\be
\bar R=\left(f\hat \alpha_r\hat p_r c+{i\hbar с f\over r} \hat
\alpha_r\hat \beta k+mc^2f_1 \hat \beta +E-U\right)R. \label{bR}
\ee
Substituting this expression into the previous equation
(\ref{barR}), we find the equation for $R$:
\be
\label{R} &&\left\{c^2(f \hat p_r)^2 +\hbar^2 c^2 k f \hat \beta
{d\over dr}\right.\left({f\over r}\right)+m^2c^4f_1^2\nonumber\\
&&\\ &&+\left.{\hbar^2 c^2 f^2 k^2\over r^2} +i\hbar c f \hat \alpha_r
{dU\over dr}-i\hbar mc^3\hat\alpha_r\hat\beta
f{df_1\over dr}-(E-U)^2\right\}R=0. \nonumber \ee

Here the separation of the space variables from those describing
the internal degrees of freedom is possible if
\be
C_1 {d\over dr}\left({f\over r}\right)={dU\over dr},\nonumber\\
\label{cc}\\ C_2 {d\over dr}\left({f\over r}\right)={df_1\over
dr}, \nonumber
\ee
where  $C_1,\ C_2$ are constants.

If (\ref{cc}) holds then equation (\ref{R}) has the form:
\be
\label{R1} &&\left\{c^2(f \hat p_r)^2 +\hbar^2 c^2 \hat \Lambda f
{d\over dr}\left({f\over r}\right)+{\hbar^2 c^2 f^2 k^2\over r^2}
+m^2c^4f_1^2-(E-U)^2\right\}R=0, \ee where the operator
\be
\hat \Lambda=k\hat \beta+ {i\over \hbar c}\hat\alpha_r
C_1-i{mc\over \hbar}\hat \alpha_r \hat \beta C_2. \label{hatlambd}
\ee

The operator  $\hat \Lambda$ does not depend on the radial
coordinate and it can easily be reduced  to the diagonal form.

Taking into account the properties of the matrices  $\hat\alpha_r$  from
(\ref{rad}) and $\hat \beta$ we choose a representation in which:
\be
\hat\beta= \left(
\begin{array}{lr}
I & 0\cr 0 & -I
\end{array}
\right), \ \ \ \ \hat \alpha_r= \left(
\begin{array}{lr}
0 & -i\cr i & 0
\end{array}
\right). \label{ab} \ee
Then operator (\ref{hatlambd}) is
\be
\hat \Lambda= \left(
\begin{array}{cc}
k & {\displaystyle {C_1\over \hbar c}+{mc\over \hbar}\, C_2}\cr
\cr {\displaystyle -{C_1\over \hbar c}+{mc\over \hbar}\, C_2} & -k
\end{array}
\right), \label{A} \ee and its eigenvalues $$\lambda=\pm
\sqrt{k^2+\left({mc\over \hbar}C_2\right)^2-\left({C_1\over \hbar
c}\right)^2}.$$

If  one works in the representation where the operator  $\hat
\Lambda$ is diagonal, our radial equation (\ref{R1}) finally
gets the following form:
\be
\label{R2} &&\left\{c^2(f \hat p_r)^2 +\hbar^2 c^2 \lambda f
{d\over dr}\left({f\over r}\right)+{\hbar^2 c^2 f^2 k^2\over r^2}
+m^2c^4f_1^2-(E-U)^2\right\}R=0.
\ee
Let us remark that as the functions  $f$, $f_1$ and $U$ are related by two conditions
(\ref{cc}), only one of them is independent, for instance, it can
be the potential energy $U$.

\section{The Kepler  problem}
\setcounter{equation}{0}

Now we consider the Kepler problem, that is the motion of the
charged particle in the Coulomb field  when the potential energy
\be
U=-{e^2\over r}, \label{U} \ee where $e^2$ is the charge square.
From  equation  (\ref{cc}) we find the deforming function
\be
f=1+\nu r, \label{f} \ee
where $\nu$ is a constant and the function
\be
f_1=1+{a\over r}, \label{f11} \ee
$a$ is a constant and
$$C_1=-e^2,\ \ \ \ \ C_2=a,$$
Then the eigenvalues of $\hat \Lambda$ are
\be
\lambda=\pm \sqrt{k^2+\left({mca\over
\hbar}\right)^2-\left({e^2\over \hbar c}\right)^2}. \label{lambda}
\ee
After a standard substitution
\be
R={\chi\over r}, \label{chir} \ee
where $\chi=\chi(r)$, the radial
equation (\ref{R2}) comes to be :
\be
\left\{-{\hbar^2 \over 2m}{d^2\over dx^2}+{\hbar^2 \over 2mr^2}\,
l^{*}(l^{*}+1) -{e^{*2}\over r}\right\}\chi=E^{*}\chi, \label{R3}
\ee
where
$$dx={dr\over f}.$$
From the latter we have
\be
x\nu=\ln (1+\nu r),\ \ \ \ \ 0\le x<\infty. \label{xnu} \ee
The values with the asterisk in equation  (\ref{R3}) are as
follows:
\be
\left\{
\begin{array}{l}
{\displaystyle l^{*}(l^{*}+1)=k^2+\left({mca \over
\hbar}\right)^2-\lambda-\left({e^2\over \hbar c}\right)^2},\cr \cr
{\displaystyle e^{*2}={E\over mc^2} e^2 -{\hbar^2 k^2 \nu\over
m}+{\hbar^2 \nu\over 2m}\lambda -{mc^2a},}\cr \cr {\displaystyle
E^{*}={E^2-m^2c^4\over 2mc^2} -{\hbar^2 k^2 \nu^2\over 2m}}.
\end{array}
\right. \label{zirk} \ee
The effective orbital quantum number
\be
l^{*}=-{1\over 2}+{1\over 2}| 2\lambda-1|=\left\{
\begin{array}{l}
\sqrt{k^2-\bar\alpha^2}-1\cr\cr \sqrt{k^2-\bar\alpha^2}
\end{array},
\right. \label{ef} \ee $$ {\bar\alpha}^2=\alpha^2-\left({mca\over
\hbar^2}\right)^2, $$
$\alpha=e^2/\hbar c$ is the fine structure
constant; here the upper value of $l^{*}$  determines   the upper
sign for $\lambda$ (\ref{lambda}) and the lower value sets the
lower sign, respectively. Thus, equation  (\ref{R3}) is split into
two independent equations for the positive and negative values of
the quantity  $\lambda$  from(\ref{lambda}). If we write the
radial coordinate  $r$  from equation (\ref{xnu}) explicitly
through  $x$ and substitute $r$ in equation(\ref{R3}) then after
simple transformations we arrive at the following equations:
\be
\left\{-{d^2\over dx^2}+{A(A-\nu/2)\over {\sinh}^2(x\nu/2)}
-{2B\over \tanh(x\nu/2)}\right\}\chi=\varepsilon\chi,
\label{chi} \ee where
\be
&&A(A-\nu/2)=\nu^2{l^{*}(l^{*}+1)\over 4},\nonumber\\ \label{ABE}
&&B={me^{*2}\nu \over 2\hbar^2}+\nu^2{l^{*}(l^{*}+1)\over 4},\\
\nonumber &&\varepsilon={2m\over
\hbar^2}\left[E^{*}-{\hbar^2\nu^2l^{*}(l^{*}+1)\over 4m}-
{e^{*2}\nu\over 2}\right].
\ee
It is well known that this equation
has the exact solution \cite{13} with the energy levels
\be
\varepsilon=-\left(A+{\nu\over 2}n_r\right)^2-{B^2\over
\left(A+{\nu n_r/ 2}\right)^2}, \label{eps} \ee
$n_r=0,1,2,\ldots$
is the radial quantum  number and bound states exist if
\be
B>A^2,\ \ \ \  A\ge 0,\ \ \ B\ge 0. \label{AB} \ee

As in our case $$A={\nu\over 2}(l^{*}+1),$$
then  from
(\ref{eps}) taking into account the notations in (\ref{zirk}) for
the energy levels  $E$ we find the following equation:
\be
&&{E^2-m^2c^4\over mc^2}= {\hbar^2 \nu^2\over 2m}
(k^2-\bar\alpha^2)-{\hbar^2\nu^2\over 4m}n^2\nonumber\\ &&+\nu e^2
{E\over mc^2}-{\nu amc^2}-{m\over \hbar^2 n^2}\left[{e^2E\over
mc^2}-mc^2a-{\hbar^2\nu\over 2m}(k^2+{\bar\alpha}^2)\right]^2,
\\
\nonumber &&n=n_r+l^{*}+1 \label{E2} \ee is the principal  quantum
number.

It is significant that the quantity  $\lambda$ drops out of this
equation  and a dependence on this quantity remains only in the
effective orbital quantum number $l^{*}$. Thus, one solution of
equation (\ref{R3}) yields the radial function
$\chi_{n_r,l^{*}}$ for $l^{*}=\sqrt{k^2-{\bar\alpha}^2}-1$ with
the energy $E=E_{n,k}$; we have the second solution for the negative
sign of the quantity  $\lambda$ in (\ref{lambda}) and  (\ref{ef}),
it equals the function $\chi_{n_r,l^{*}+1}$ with the eigenvalue of energy
$E_{n+1,k}$. In the nonrelativistic case, the first solution gives
$l^{*}=l=0,1,2,\ldots$, and the second one  $l^{*}=l=1,2,\ldots$
where  $l$ is the usual orbital quantum number. Thus the energy
levels for the two solutions coincide with the exception of the
ground state. Here we have the so-called super-symmetry. The Dirac
equation  (\ref{Dir3}) for the Coulomb potential with the
deforming functions  $f$ and $f_1$ satisfying  conditions  (\ref{cc})
reveals supersymmetry. But this issue calls for a
separate study. Solving equation (\ref{E2}) for
$E=E_{n,k}$  we finally find
\be
E&=&{\nu e^2\over 2} {(n^2+k^2+\bar\alpha^2)\over
n^2+\alpha^2}+\left({mc\over \hbar}\right)^2{e^2a\over
n^2+\alpha^2}\nonumber
\\ \label{E}
&+&{mc^2\over {1+\alpha^2/n^2}} \left\{1+{{\bar\alpha}^2\over
n^2}+\left({\nu e^2\over 2mc^2}\right)^2
\left(1+{k^2+\bar\alpha^2\over n^2}\right)
\left(1+a\nu+{k^2+\bar\alpha^2\over n^2}\right)\right.
\\
&+&\left.\left({\hbar\nu\over 2mc}\right)^2 \left(1+{\alpha^2\over
n^2}\right)
\left[2(k^2-\bar\alpha^2)-n^2-{(k^2+\bar\alpha^2)^2\over
n^2}\right] \right\}^{1/2},\nonumber \ee

The condition for the existence of bound states follows from
(\ref{AB}):
\be
{E\over mc^2}e^2>{\hbar^2 \nu\over m}k^2+mc^2a. \label{isn} \ee

The initial function  $\Psi$ contained in equation
(\ref{Dir1}) is found from (\ref{peretv}), (\ref{psir}),
(\ref{bR}) and (\ref{chir}):
\be
\Psi=f^{-1/2}Y\left(f\hat \alpha_r\hat p_r c+{i\hbar с f\over r}
\hat \alpha_r\hat \beta k+mc^2f_1 \hat \beta +E-U\right){\chi\over
r}, \label{psiY} \ee
where  $\chi$   is the matrix-column with
the elements $\chi_{n_r,l^{*}}$ and $\chi_{n_r,l^{*}+1}$.

Formulae  (\ref{E})--(\ref{psiY}) provide the exact solution
of the Kepler problem in the Dirac theory with Heisenberg algebra
that is deformed by function  (\ref{f}) with the position
dependent particle mass  in accordance with (\ref{mz}), (\ref{f11}).

\section{Discussion of the results}
\setcounter{equation}{0}

If in  (\ref{E}) we put $\nu=0$, i.e. we remove deformation, the
energy levels for the Dirac charged particle whose mass is position-dependent are
obtained:
\be
E={mc^2\over 1+\alpha^2/n^2} \left({m e^2a\over \hbar^2
n^2}+\sqrt{1+{\bar\alpha^2\over n^2}}\right), \label{51} \ee
and
$$a<{e^2\over mc^2}$$
that follows from (\ref{isn}).
This result was originally discovered in \cite{Soff} and
reproduced in \cite{14} by a different technique.

The nonrelativistic limit, $c\to \infty$, for expression
(\ref{E}) was found. We assume that the function  $f_1$ satisfies
condition (\ref{f1}), otherwise said, we believe that the
dependence  of the particle mass on its coordinates makes its own contributions into
the nonrelativistic limit. It means that taking into
account the explicit form of the function  $f_1$ (\ref{f11}) the
parameter  $a\sim 1/c^2$. That is why we take
$$ a={e^2\over mc^2}\, \bar a,$$
where   $\bar a$ is a dimensionless constant. In this
case the nonrelativistic limit for the energy  $E$ is as
follows:
\be
E'&=&E-mc^2=-{m\over 2\hbar^2 n^2}\left(e^2-{\hbar^2 \nu\over
2m}k^2\right)^2-{\hbar^2 \nu^2\over 8m} n^2 +{\nu\over 2}
\left(e^2+{\hbar^2 \nu\over 2m}k^2\right)\nonumber\\ &+&
{me^4\over 2\hbar^2 n^2}\bar a (2-\bar a), \label{EE} \ee
and in
accordance with (\ref{isn}) the energy  spectrum is limited.
\be
e^2> {\hbar^2 \nu\over m}k^2+{e^2 \bar a}. \label{e2} \ee
It is
interesting to compare expressions  (\ref{EE}) and (\ref{e2})
when  $\bar a=0$ with the results in  \cite{15} where  the
Schr\"odinger equation for a particle in the  Coulomb field
with the deforming function (\ref{f}), in our notation, was solved:
\be
E'_{\rm QT}=-{m\over 2\hbar^2 n^2}\left(e^2-{\hbar^2 \nu\over
2m}\left[l(l+1)+1\right]\right)^2-{\hbar^2 \nu^2\over 8m} n^2
+{\nu\over 2} \left(e^2+{\hbar^2 \nu\over
2m}\left[l(l+1)+1\right]\right), \label{QT} \ee
with the condition that
$$ e^2> {\hbar^2 \nu\over 2m}\left[(l+1)(2l+1)+1\right].$$

The difference of this expression from formula
 (\ref{EE})
is explained by the fact that the authors of  \cite{15}
disregarded the deformational spin-orbital interaction  $\Delta
U$ which arises naturally  in our treatment in the nonrelativistic
limit from the Dirac equation. These authors started from
the Schr\"odinger equation at once. If the said interaction is not taken into
account, we must deduce the contribution from
\be
\Delta U= {\nu^2\over m}(\hat{\bf S}\hat{\bf L})+ {\nu\over
m}(\hat{\bf S}\hat{\bf L}){1\over r},
\label{delU} \ee
which follows from  (\ref{dU}) and (\ref{f}). As the eigenvalue of the
operator  $(\hat{\bf S}\hat{\bf L})=(\hat{\bf J}^2-\hat{\bf
L}^2-\hat{\bf S}^2)/2$ equals
$\hbar^2[j(j+1)-l(l+1)-3/4]/2=\hbar^2[(j+1/2)^2-l(l+1)-1]/2=
\hbar^2[k^2-l(l+1)-1]/2$ this contribution can be easily taken
into account. Indeed, in order to remove  the contribution of
$\Delta U$  from our result it is necessary to deduce from the
energy  $E'$  the contribution of the first term in
  (\ref{delU}) equaling $\hbar^2\nu^2[k^2-l(l+1)-1]/2m$; the second term in (\ref{delU})
should be united with the Coulomb potential (\ref{U}) by the
substitution: $e^2\to e^2+\hbar^2\nu[k^2-l(l+1)-1]/2m$.
Consequently, from  (\ref{EE}) we arrive at  expression
(\ref{QT}). Besides, we must put $j=l+1/2$ in  condition
(\ref{e2}) limiting the spectrum in expression
(\ref{operat}) for the quantum number $k$. In other words, we
should take a higher value of  $k^2=(l+1)^2$.

Now we give  the next after the zeroth approximation (\ref{EE}) term of the development
of energy $E^{(1)}$ by the degrees $1/c^2$. We represent $E^{(1)}$ as a sum of three
terms:
\be
E^{(1)}=\Delta_1 E^{(1)}+\Delta_2 E^{(1)}+\Delta_3 E^{(1)}. \ee
The correction does not depend on the parameter  $\nu$
\be
\Delta_1 E^{(1)}=-{me^4\over 2\hbar^2}{\alpha^2\over n^4} (1-\bar
a)^3\left[{n\over |k|}(1+\bar a)-{3\over 4} (1+\bar a /3)\right].
\ee
At $\bar a=0$ it transforms into the well-known Sommerfeld formula.
The correction
\be
\Delta_2 E^{(1)}=-\left({\hbar \nu\over
8mc}\right)^2{\hbar^2\nu^2\over 2m} {(n^2-k^2)^4\over n^4}, \ee
is brought about only by deformation.
The cross term
\be
\Delta_3 E^{(1)}&=&{\nu e^2\over 2} {\alpha^2\over n^4}
\left[(1-\bar a^2)n|k|-k^2-n^2\bar a^2\right]+{\hbar^2 \nu^2\over
8m}{\alpha^2\over n^4}\Bigg\{(n^2+k^2)^2\\ &+&(1-\bar
a^2)\left[2k^4-{3\over 2}(n^2+k^2)^2+{n\over |k|} (n^4-k^4)
\right]\Bigg\}. \nonumber \ee

The obtained results are of general interest. They can also be
useful for the study of the energy spectrum of nanoheterosystems
when the electrons mass is position
dependent and also whenever it is important to  take into account
relativistic effects, in particular those of
spin-orbital interaction.

In the end, let us mention that the question of the application of deformed commutational relations
in Kepler relativistic problem remains open.
As non-deformed Kepler problem is Lorenz-invariant the question
arises whether this property will be preserved in the
deformed space. Though it is well-known that the quantum
space-time with deformed Heisenberg algebra can be
Lorenz-invariant \cite{11}, it is obvious that in our case the
problem is not like that. A similar controversy  was found in
\cite{16} that studied the Dirac oscillator with deformed commutational
relations  leading to the existance of the minimal length of
space. However, this problem calls for a more detailed investigation to be
suggested in my next paper.

The author is very grateful to V.~M.~Tkachuk for fruitful
discussions.

\end{document}